\documentclass[a4paper]{cas-dc}

\usepackage[authoryear]{natbib}
\usepackage{hyperref}
\usepackage{ulem}
\usepackage{xcolor}  
\definecolor{Revision}{RGB}{0, 0, 0}
\def\tsc#1{\csdef{#1}{\textsc{\lowercase{#1}}\xspace}}
\tsc{WGM}
\tsc{QE}
\tsc{EP}
\tsc{PMS}
\tsc{BEC}
\tsc{DE}

\begin{document}
\let\WriteBookmarks\relax
\def\floatpagepagefraction{1}
\def\textpagefraction{.001}
\let\printorcid\relax
\shorttitle{A topology-preserving three-stage framework for fully-connected coronary artery extraction}
\shortauthors{Yuehui Qiu et~al.}

\title [mode = title]{A topology-preserving three-stage framework for fully-connected coronary artery extraction}                    



\author[1]{Yuehui Qiu}[type=editor]
\credit{Writing – original draft, Methodology, 
Data curation, Conceptualization}
\fnmark[1]
\affiliation[1]{organization={Center for Digital Media Computing, School of Film, School of Informatics, Xiamen University},
                city={Xiamen},
                postcode={361005}, 
                country={China}}

\author[2]{Dandan Shan}[type=editor]
\credit{Writing – original draft, Methodology, Data curation, 
 Conceptualization}
\fnmark[1]
\affiliation[2]{organization={Institute of Artificial Intelligence, Xiamen University},
                city={Xiamen},
                postcode={361005}, 
                country={China}}

\author[3]{Yining Wang}
\credit{Writing – review \&
editing}
\affiliation[3]{organization={Peking Union Medical College Hospital},city={Beijing},postcode={100006}, 
                country={China}}
                
\author[4]{Pei Dong}
\credit{Writing – review \&
editing}
\affiliation[4]{organization={Shanghai United Imaging Intelligence Co., Ltd.}, city={Shanghai},postcode={200232},  country={China}}

\author[4]{Dijia Wu} 
\credit{Writing – review \&
editing}

\author[5]{Xinnian Yang}
\credit{Writing – review \&
editing}
\affiliation[5]{organization={City University of Hong Kong}, city={Hong Kong},postcode={999077},  country={China}}

\author[1,2]{Qingqi Hong}
\credit{Writing – review
\& editing, Supervision, Project administration}
\cormark[1]
\ead{hongqq@xmu.edu.cn}

\author[4,6,7]{Dinggang Shen}
\credit{Writing – review
\& editing, Supervision, Project administration}
\cormark[1]
\ead{dinggang.shen@gmail.com}
\affiliation[6]{organization={School of Biomedical Engineering \& State Key Laboratory of Advanced Medical Materials and Devices, ShanghaiTech University},city={Shanghai},
                postcode={201210}, country={China}}
\affiliation[7]{organization={Shanghai Clinical Research and Trial Center},city={Shanghai},postcode={200231}, country={China}}
\cortext[cor1]{Corresponding authors: Qingqi Hong and Dinggang Shen.}
\fntext[fn1]{Yuehui Qiu and Dandan Shan have equal contribution to this work.}



\begin{abstract}
Coronary artery extraction is a crucial prerequisite for computer-aided diagnosis of coronary artery disease. Accurately extracting the complete coronary tree remains challenging due to several factors, including presence of thin distal vessels, tortuous topological structures, and insufficient contrast. These issues often result in over-segmentation and under-segmentation in current segmentation methods. To address these challenges, we propose a topology-preserving three-stage framework for fully-connected coronary artery extraction. This framework includes vessel segmentation, centerline reconnection, and missing vessel reconstruction. First, we introduce a new centerline enhanced loss in the segmentation process. Second, for the broken vessel segments, we further propose a regularized walk algorithm to integrate distance, probabilities predicted by a centerline classifier, and directional cosine similarity, for reconnecting the centerlines. Third, we apply implicit neural representation and implicit modeling, to reconstruct the geometric model of the missing vessels. 
{Experimental results show that our proposed framework outperforms existing methods, achieving Dice scores of 88.53\% and 85.07\%, with Hausdorff Distances (HD) of 1.07mm and 1.63mm on ASOCA and PDSCA datasets, respectively.} Code will be available at \href{https://github.com/YH-Qiu/CorSegRec}{https://github.com/YH-Qiu/CorSegRec}.

\end{abstract}




\maketitle

\section{Introduction}

\label{sec:introduction}
Coronary artery disease, a severe cardiovascular condition, is recognized as one of the leading causes of death worldwide. In diagnosing and treating this disease, coronary artery segmentation technology plays a crucial role. However, traditional segmentation methods \citep{lesage2016adaptive} require manual parameter 
\begin{figure}[ht]
   \centering
\includegraphics[width=0.95\linewidth]{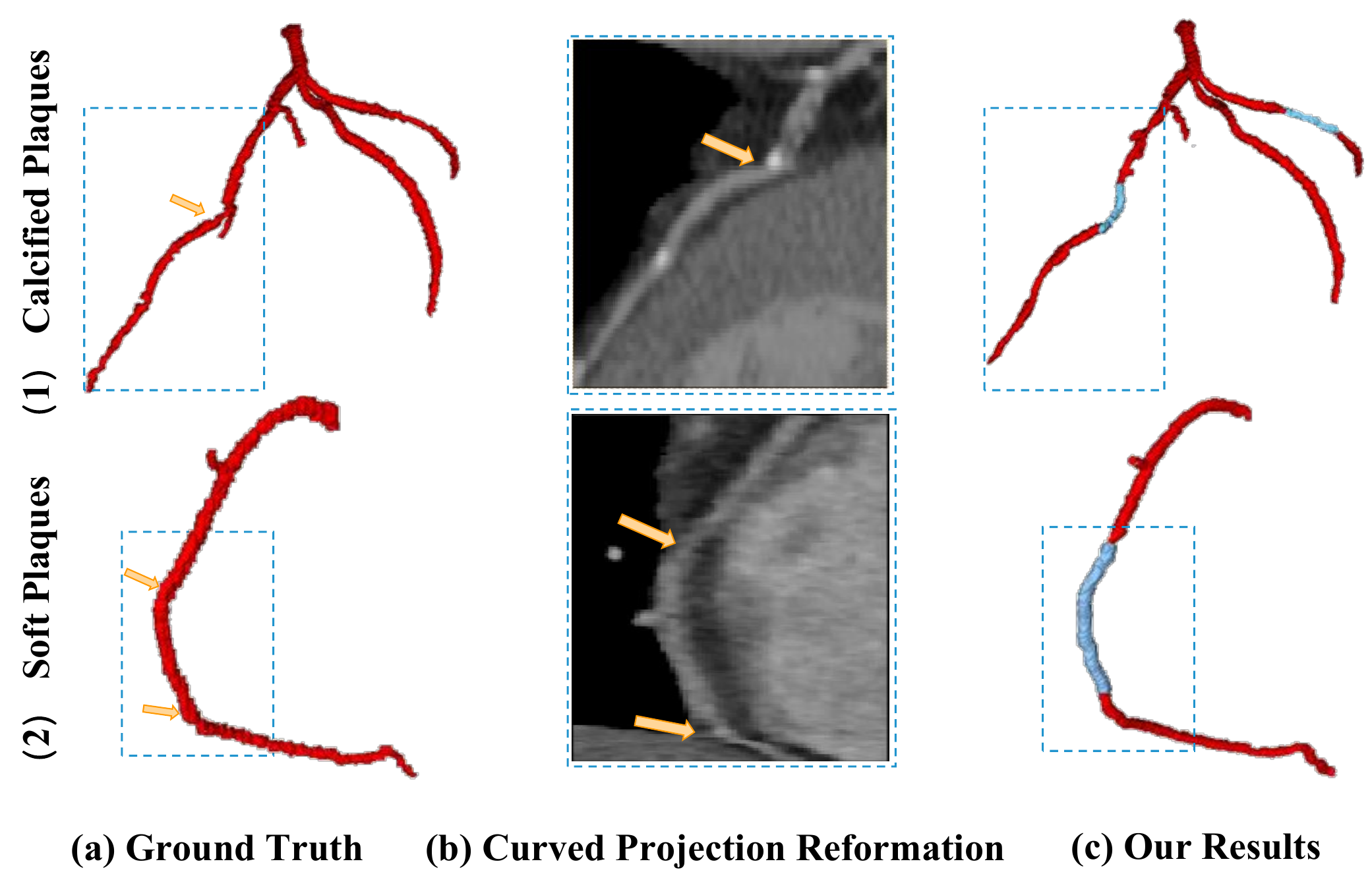}
  \caption{
  The reconstruction of the missed vessel models for disconnected branches. (1) The yellow arrow points to the disconnected region. Due to presence of calcified plaque, the corresponding location in the ground truth shows significant luminal narrowing, and our initial segmentation also shows discontinuity at the corresponding location. (2) Two yellow arrows point to two ends of large disconnected segment on the right coronary artery. At the starting end, the vessel appears "unclear" due to presence of soft plaque, while, near the terminal end, the vessel shows mild narrowing.
  }
  \label{Fig_Intro}
  \setlength{\belowcaptionskip}{-1cm}
  \vspace{-6mm}
\end{figure}
tuning to adapt to different datasets, which is time-consuming and labor-intensive. With the successful application of deep learning, deep convolutional neural networks \citep{duan2023eca}
 have been widely employed in the field of medical image segmentation, including coronary computed tomography angiography (CCTA). Generally, these deep learning methods adopt end-to-end schemes, and achieve higher levels of automation and accuracy by transforming the segmentation problem into a voxel-wise classification problem. 
 Among them, nnU-Net \citep{isensee2021nnu} has attracted considerable attention due to its flexible network design, powerful data augmentation, scalable pre-trained models, and excellent performance, making it an important tool in the field of medical image segmentation. To address the segmentation problem of tubular structures, a series of specially designed network architectures have also emerged. For example, Mou \textit{et al}. \citep{mou2021cs2} proposed CS$^{2}$-Net, which incorporated channel attention and spatial attention mechanisms to capture long-range dependency and further designed specific convolutional kernels to more effectively segment curvilinear  structures. Additionally, Wolterink \textit{et al}. \citep{wolterink2019graph} 
 applied graph convolutional networks to coronary artery segmentation, optimizing 
 the points on the luminal surface to generate smooth surface contours.

However, due to extremely small proportion of coronary arteries in the entire CCTA volume, the low contrast with artifacts and lesions as well as the distraction from neighboring  tubular structures, such as coronary veins,  the coronary segmentation remains a challenging problem. In addition, as shown in Fig. \ref{Fig_Intro} (b), the presence of plaques causing the stenosis can lead to discontinuity in vessel segmentation. Thereby, improving accuracy and connectivity of coronary artery segmentation continues to be a significant focus of ongoing research.
Qiu \textit{et al}. \citep{qiu2024deep} employed multi-scale dilated convolution techniques to capture a broader range of contextual information for alleviating vessel rupture, but did not take into account the topology of the coronary arteries.
He \textit{et al}. \citep{he2023automated} introduced graph attention networks into CNN to assist in learning the global geometric information of coronary arteries. However, it is less effective when capturing geometric structures of small vessels.



To address the aforementioned challenges, in our previous work \citep{qiu2023corsegrec}, we developed a coronary artery extraction framework comprising three stages: vessel segmentation, vessel reconnection, and vessel reconstruction. By first performing coarse segmentation of the vessels and then segmenting contours of disconnected vessels based on vascular topological structure, our method can effectively reconnect interrupted centerlines and fill in missing lumens. This approach significantly enhances the integrity and connectivity of the coronary artery tree.
However, our previous work have two main limitations: (1) The DPC (Distance-Probability-Cosine) walk algorithm does not account for situations where the predicted continuous centerline does not include the branch containing the "disconnected" centerline during reconnection. (2) The model-based level set method generally provides coarse segmentation of narrow lumens, resulting in significant gap between the segmented results and the ground truth, in addition to high memory cost.


Therefore, in this study, we extend our previous work to address these limitations. We have improved the DPC walk algorithm to address cases where the predicted continuous centerline does not include the branch containing the "disconnected" centerline, allowing it to handle more complex repairment scenarios. We also design a two-level neighborhood set to handle long-distance reconnection processes. Furthermore, we adopt a contour extraction algorithm based on implicit neural representation (INR) \citep{molaei2023implicit} to extend the signed distance function (SDF) based zero-level set representation of vessel contours from the initial contour dynamic points of 2D sections. This approach can better address voxel grayscale inhomogeneity, reduce memory overhead, improve processing speed, and enhance segmentation accuracy.

The main contributions of this study can be summarized as follows:

\begin{itemize}
\item  We develop a fully-connected coronary artery extraction framework consisting of three stages: vascular segmentation, centerline reconnection, and missing vessel reconstruction, to address the issues of over-segmentation and under-segmentation in current coronary artery segmentation methods.
\item To handle more complex cases of disconnected vessels, we improve the DPC (Distance-Probability-Cosine) traversal strategy, enhancing the overall segmentation accuracy and connectivity.  
\item  To rapidly generate smooth luminal regions in repaired areas, we design a segmentation model based on INR to segment the "unclear" lumen, complemented by skeleton-based vascular reconstruction techniques, achieving fully-connected coronary artery extraction (Fig. \ref{Fig_Intro} (c)).
\end{itemize}

\section{Related work}
\subsection{Coronary artery segmentation methods}
Coronary artery segmentation is a crucial area in medical image processing, aiming to accurately extract geometric structural information of blood vessels from coronary artery imaging data to assist physicians in diagnosis and treatment planning. Traditional methods mainly rely on mathematical models and manually designed rules 
\citep{lesage2009review,ma2020coronary,mohr2012accurate}, which are time-consuming and labor-intensive. In recent years, deep learning techniques, including models like U-Net \citep{ronneberger2015u} {and ResUNet \citep{kerfoot2019left}}, have made significant advance in coronary artery segmentation, effectively extracting features and achieving accurate segmentation through end-to-end learning. For instance,
{Zhu \textit{et al}. \citep{zhu2021coronary} introduced a PSPNet-based multi-scale CNN model for segmentation of coronary angiography images. Song \textit{et al}. \citep{song2022automatic} proposed a two-stage network, consisting of a classification module and a segmentation module, for coronary artery segmentation.}
Jiang \textit{et al}. \citep{jiang2024ori} used multi-task learning to generate coarse segmentation, radius, and orientation, and employed an orientation-guided tracking method to iteratively reconstruct coronary arteries, thereby improving segmentation performance.
Dong \textit{et al}. \citep{dong2023novel} used attention-guided feature fusion between adjacent layers in the encoding and decoding process to enhance the generalization capability of coronary artery segmentation. Zhang \textit{et al}. \citep{10265156} proposed a two-stage anatomy and topology  preserving framework, combining an anatomical dependency encoding module and a hierarchical topology learning module for coarse-to-fine segmentation.

\subsection{Vascular reconnection methods}
Coronary artery trees obtained by learning methods \citep{zeng2023imagecas} or partial traditional vessel segmentation methods \citep{ma2020coronary} are usually not fully connected, with some "disconnected" vessels. They may include tubular structures that are not part of the coronary artery, or they may be disconnected from the coronary tree due to stenosis or the presence of plaques. Many methods have been used to divide blood vessels, but the reconnection of disconnecting blood vessels has received less attention.
Vascular reconnection aims to restore cohesive vascular tree structures by extracting centerlines or segmenting lumens in local images using tracking or growth techniques. 
{
For example, Naeem \textit{et al}. \citep{Naeem2024MICCAI} proposed a recurrent DETR based model to track topologically correct centerlines of tubular tree objects in 3D volumes; and, recently, Ye \textit{et al}. \citep{Ye2025VSR} developed a novel network with graph clustering for vessel-like structure rehabilitation.
}
In the task of coronary artery CTA centerline extraction, Yang \textit{et al}. \citep{yang2012automatic} used a wave propagation algorithm to reconnect the disconnected branches to the initial tree in a rectangular search area. 
Li \textit{et al}. \citep{li2022deep} proposed a vascular tracking network to grow local coronary arteries to make up for gaps and discontinuities in segmentation results that might be caused by artifacts or diseases. Other studies have looked at non-CTA or non-coronary artery repair. For example, Han \textit{et al}. \citep{han2022reconnection}  used convolutional neural network considering local details and local geometric prior to reconnect the coronary artery fragments in X-ray angiography images, with increase of Dice coefficient for 0.33. While all these studies have integrated local vascular details and geometric priors in vascular reconnection, there is currently no strategy tailored specifically for coronary reconnection in 3D CTA that is capable of effectively managing complex disconnections and achieving a comprehensive coronary tree.

\subsection{Implicit neural representation in medical imaging}
Implicit neural representation (INR) has found widespread application in medical image reconstruction and segmentation, effectively addressing the inefficiencies of deep learning in handling complex topological structures. In recent years, they have been primarily used for resolution enhancement and missing-information synthesis. NeRP \citep{shen2022nerp} is an implicit neural representation learning method used for reconstructing computational images from sparsely sampled measurements. It leverages image priors and physical information of sparsely sampled measurements to generate representations of unknown subjects, exhibiting broad applicability and robustness across different imaging modalities while capturing subtle yet significant image changes. Reed \textit{et al}. \citep{reed2021dynamic} utilized implicit neural representations and a novel parametric motion field deformation for limited-view 4D-CT reconstruction, employing a differentiable analysis-by-synthesis approach to optimize reconstruction without requiring training data and demonstrating robust reconstruction of scenes with deformable and periodic motion. Barrowclough \textit{et al}. \citep{barrowclough2021binary} proposed an image segmentation method that combines implicit spline representations with deep convolutional neural networks, for achieving segmentation of congenital heart disease by pre- dicting control points of bivariate spline functions. Gu \textit{et al}. \citep{gu2023retinal} utilized implicit neural representations to enhance the resolution of original retinal images and employed a texture enhancement module to enhance vessel details, improving the segmentation of small and discontinuous vessels.
{
Recently, Fu \textit{et al}. \citep{Fu2024MICCAI} proposed a 3D Gaussian representation technique for coronary artery reconstruction from ultra-sparse X-ray projections.
}
\section{Methods}
\subsection{ Overview of CorSegRec }
\begin{figure*}[htbp]
\setlength{\abovecaptionskip}{-1mm}
\centering 
\includegraphics[width=0.95\textwidth]{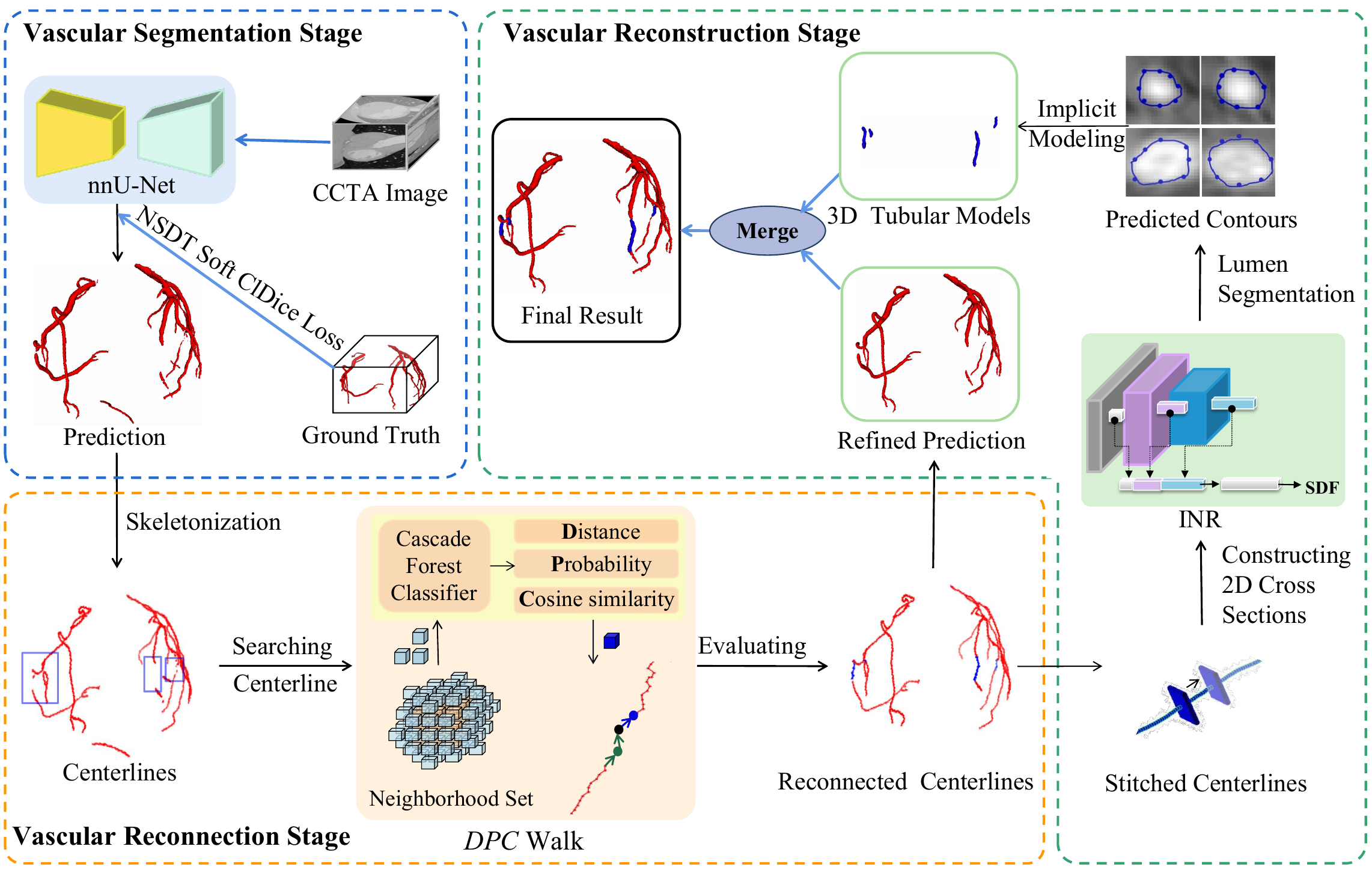}
\caption{Overview of the proposed framework. CorSegRec consists of three stages: vascular segmentation, vascular reconnection, and vascular reconstruction.}
\label{fig:corsegrec}
\end{figure*}

As shown in Fig. \ref{fig:corsegrec}, we construct a fully-connected coronary artery extraction scheme named CorSegRec (\textbf{Cor}onary \textbf{Seg}mentation \textbf{Rec}onnection), which includes three stages: vascular segmentation stage, vascular reconnection stage, and vascular reconstruction stage.
In the vascular segmentation stage, we utilize the advanced medical image segmentation framework nnU-Net to produce the initial segmentation outcomes of coronary arteries. Given the significance of topological predictive capability in influencing the potential for vascular reconnection, we introduce a novel centerline-enhanced loss function for coronary artery structure extraction in Stage 1. This function, termed NSDT (Normalized Skeleton Distance Transform) Soft-clDice (Soft-CenterlineDice) loss, is designed to augment the segmentation model’s ability to extract coronary artery topology.

In the vascular reconnection stage, the initial segmentation results are refined to extract and segment the centerline of the vessel. Target vessels for reconnecting disconnected vessels are identified based on the distances between disconnected vessels and the vascular connecting tree, as well as the directions of the ends of each connected vessel. Subsequently, a centerline reconnection algorithm, termed DPC (Distance-Probability-Cosine) walk algorithm, is proposed to locally search the centerline iteratively. The integrity of the repaired centerline sequence is evaluated for reliability before consid-ering it as a successful reconnection and proceeding to the next stage. Any segments that fail to reconnect successfully are eliminated from the prediction mask.

In the vascular reconstruction stage, 2D cross-sections are generated along the centerline of the stitched vessel. Subsequently, to enhance segmentation efficiency and accuracy, INR is utilized to efficiently segment the coronary lumen. By extruding the lumen profiles into 3D vessel models, the ultimate segmentation results are attained through the integration of the reconstructed vessel models with the prediction mask of the problematic vessel segment.
\subsection{Vascular segmentation stage}
In the segmentation stage, to better detect the fine vessels in the distal coronary arteries, we introduce NSDT Soft-clDice for enhancing the weights of vascular centerlines by computing the normalized distance from mask foreground voxels to the vascular skeleton.
\begin{equation}
\label{eq:ldsc}
\resizebox{0.75\linewidth}{!}{
    $ L_{dscl} = 1 - 2 \times \frac{Tprec_{c}^{*}(S_P, V_L) \times Tsen_{s}^{*}(S_L, V_P)}{Tprec_{c}^{*}(S_P, V_L) + Tsen_{s}^{*}(S_L, V_P)} $
}
\end{equation}

The structure is similar to that in \citep{shit2021cldice}, but new calculations for topology precision and topology sensitivity are defined:
\begin{equation}
\label{eq:tprec}
\resizebox{0.75\linewidth}{!}{
    $ Tprec_{c}^{*}(S_P, V_L) = \frac{\left| (S_P \circ NSDT_P \circ V_P) \circ NSDT_L \right|}{\left| (S_P \circ NSDT_P) \circ (S_P \circ NSDT_P) \right|} $
}
\end{equation}
\begin{equation}
\label{eq:tsens}
\resizebox{0.75\linewidth}{!}{
    $ Tsen_{s}^{*}(S_L, V_P) = \frac{\left| (S_L \circ NSDT_L) \circ (NSDT_P \circ V_P) \right|}{\left| (S_L \circ NSDT_L) \circ (S_L \circ NSDT_L) \right|} $
}
\end{equation}
where $V_L$ and $V_P$ represent label and  probability prediction, respectively. $S_L$ and $S_P$ represent the skeletons extracted using the soft-skeletonize method for $V_L$ and $V_P$, respectively.
{$\circ$ denotes the Hadamard product. Eq. \eqref{eq:tprec} computes the fraction of $S_P$ that lies within $V_L$, called topology precision, and Eq. \eqref{eq:tsens} computes vice-a-versa topology sensitivity. In both equations, we first weight skeletons and volumes by NSDT operations respectively, and then calculate their intersections.}

The NSDT is defined as:
\begin{equation}
\label{eq:nsdtl}
\resizebox{0.75\linewidth}{!}{
    $ NSDT_L = \left\{ \begin{array}{ll}
    \mathop {\inf }\limits_{y \in (S_L)_{in}} \frac{R}{\|x - y\|_2 + 1}, & x \in (V_L)_{in} \\
    0, & \text{otherwise}
    \end{array} \right. $
}
\end{equation}
where the Euclidean distance from the foreground voxel to the skeleton voxel of the mask is calculated and then multiplied by the amplification factor R after normalization. After the distance transformation, the weight of all the vascular skeleton increases to R, and the weight gradually decreases from the vascular skeleton to the vascular wall. On non-foreground voxels, the weight is 0.
{With NSDT Soft-ClDice, the network will try to produce more accurate topology and maximize both precision (removing more unnecessary segments) and sensitivity (detecting more difficult-to-divide vessels).}
The basic segmentation network used in the vascular segmentation stage is nnU-Net. 

In order to maintain the stability of the training process, we use Dice and the proposed NSDT soft-clDice as a joint loss function, defined as:
\begin{equation}
\label{eq:ldc_dscl}
\resizebox{0.6\linewidth}{!}{
    $ { L}_{dc\& dscl} = (1 - \alpha ){ L}_{dc} + \alpha { L}_{dscl} $
}
\end{equation}
where $\alpha$ is the equilibrium coefficient, and set to 0.5.

\subsection{Vascular reconnection stage}
Before vascular reconnection, the initial segmentation is skeletonized to extract the centerline branches of the two largest connected components, denoted as $V_{i}$. For other disconnected branches, their centerline branches are extracted and denoted as $CL_{j}$. Neighborhood search is performed using $11 \times 11 \times 11$ patches to detect the opening points.

\subsubsection{Candidate branch selection}
\begin{figure}[ht]
  \centering
  \includegraphics[width=\linewidth]{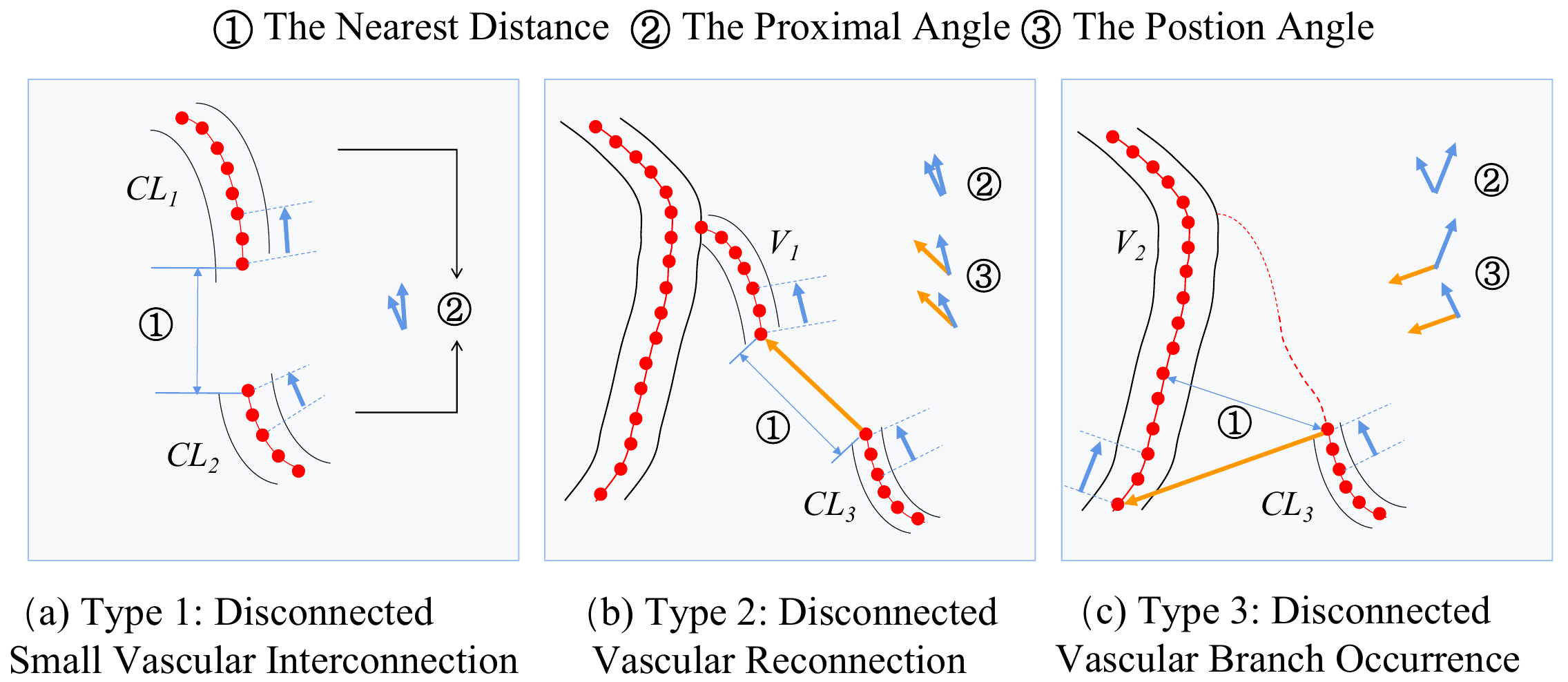}
  \caption{
  Candidate branch filtering. 
  }
  \label{candidate}
  \setlength{\belowcaptionskip}{-1cm}
  \vspace{-3mm}
\end{figure}

Due to complexity of the coronary artery’s topology, we sequentially perform three types of reconnection on all input centerline branches to obtain the reconnected centerline. First, small vessel interconnections form a new disconnected centerline, avoiding path overlap issues. Second, disconnected vessel connections consider each disconnected vessel centerline for connection to the end of a connected centerline. Finally, if the predicted connected centerline does not contain the branch of the disconnected centerline, the reconnection becomes a branch occurrence process.
As shown in Fig. \ref{candidate}, the selection of candidate branches for the "disconnected" centerline is primarily based on distance and direction. 
The nearest distance is defined as the minimum distance between all points of the two centerlines. The proximal angle is defined as the angle between the direction vectors of \textit{the "head" of the disconnected vessel} and \textit{the "tail" of the candidate vessel}, calculated from three consecutive centerline points. The positional angle is defined as the angle between the connection vectors of the two vessels and their respective direction vectors.
{The thresholds for distance and angles are empirically set to strike a balance between minimizing total reconnection time expenditure and ensuring comprehensive inclusion of all essential vascular pairs requiring reconnection. }For Type 1, the nearest distance is set to less than 60, and the proximal angle must be less than 120°. For Type 2, the nearest distance is set to less than 80, the angle between proximal endpoints is less than 120°, and the sum of the cosine of positional angles between the two positions must be greater than the minimum of 1.6 and twice the cosine of the angle between the proximal endpoints. Type 3 handles cases where branches are rare in total connection scenarios, focusing only on closely located and long-disconnected centerlines.
If the length exceeds 10, the nearest distance is set to 20. If the disconnected vessel's length exceeds 50, the nearest distance is set to 80. Proximal angle and positional angle settings are reversed compared to Type 2. For all "disconnected" centerlines, no more than two candidate branches meeting the criteria are selected in order of filtering. Once a pair of "disconnected" centerlines and candidate branches are obtained, the reconnection process begins.

\subsubsection{DPC walk}
To connect $CL_{j}$ to candidate branch $V_{i}$, we design the DPC walk algorithm. We use a distance metric (D) to guide the walk directly towards the endpoint of the candidate branch. {Additionally, we train a centerline classification model (P) based on Deep Forest \citep{zhou2019forest} to predict the probability of vascular centerlines. Deep Forest \citep{zhou2019forest} is a tree-based ensemble deep learning framework with high training efficiency and automatic determination of model complexity. }The cosine similarity of vectors (C) is used to measure the value of historical centerlines. 
Assume the current position of the walker is point A. To avoid inaccurate local decisions influenced by nearby points, we employ a two-level neighborhood set with a side length of 5 centered at point A, restricting neighbors within a distance of 2 to 3 during reconnection processes with larger nearest distances. 
The calculations of D, P, C are defined:
\begin{equation}
\label{eq_D}
\resizebox{0.4\linewidth}{!}{
    $D({A_k}) =  - {\left\| {{A_k} - {p_m}} \right\|_2}$
    }
\end{equation}
\begin{equation}
\label{eq_P}
\resizebox{0.9\linewidth}{!}{
    $P({A_k}) = {{CFC}}(flat(maxpool(\mathop {pat}\limits_{big} ({A_k}))) + flat(\mathop {pat}\limits_{small} ({A_k})))$ }
\end{equation}
\begin{equation}
\label{eq_C}
\resizebox{0.6\linewidth}{!}{
    $ C({A_k}) = \cos ({{\bf{o}}_k},{{\bf{o}}_{ - 1}}) + \cos ({{\bf{o}}_k},{{\bf{o}}_{ - 2}}) $
}
\end{equation}
where $A_{k}$ represents a point in the neighborhood set of A. $p_{m}$ represents the candidate branch endpoint. D($A_{k}$) represents the negative inverse distance from $A_{k}$ to  $p_{m}$. $o_{k}$ represents the offset vector of $A_{k}$ relative to A. C($A_{k}$) represents the cosine sum of $o_{k}$ and the offset vectors ($o_{-1}$, $o_{-2}$) from the previous two movements. P($A_{k}$) represents the probability that the point $A_{k}$ is located on the centerline.
{Its training and prediction are carried out by Cascade Forest Classifier (CFC), which is implementation of the Deep Forest \citep{zhou2019forest} for classification. Specifically, we construct two image patches ($pat$) of different sizes centered at $A_{k}$, resize them to a uniform dimension through pooling operations, and then flatten and concatenate them into a one-dimensional vector to feed into CFC.} During reconnection, P occasionally drops to significantly low levels, disrupting equilibrium with D and C. Therefore, max-min normalization is applied to the P values of all neighbors of A to obtain $P_N$, ensuring that the probability values have relatively stable discrimination:
\begin{eqnarray}
    & {P_N}({A_k}) = \frac{{P({A_k}) - \mathop {\min }\limits_{{A_t} \in \Theta } P({A_t})}}{{\mathop {\max }\limits_{{A_t} \in \Theta } P({A_t}) - \mathop {\min }\limits_{{A_t} \in \Theta } P({A_t})}}  
\end{eqnarray}
where $\Theta$ represents the set of neighbors. 

After the weighted sum of these three indicators, the comprehensive index DPC for evaluating the possibility of migration is obtained, and its calculation is defined as follows:
\begin{eqnarray}
    \resizebox{0.95\linewidth}{!}{$DPC({A_k}) = \left\{ \begin{array}{*{20}{c}}
    {D({A_k}) + \omega {P_N}({A_k}) + C({A_k}),}&{\cos ({{\bf{o}}_{ - 1}},{{\bf{o}}_{ - 2}}) \le \frac{1}{2}}\\
    {D({A_k}) + \omega {P_N}({A_k}),}&{otherwise}
    \end{array} \right.$}
\end{eqnarray}
where $\omega$ is set to 5. When the direction of the offset vector in the first two steps of history is relatively close, the C term is not involved in the calculation, to avoid forming a straight line during the walk.

When searching for the next centerline point, nodes with angles exceeding $90^\circ$ relative to historical directions are filtered out. Already reconnected centerline points are also filtered. Then, the remaining neighbor nodes become candidates and enter the DPC calculation process. The candidate with the highest DPC value becomes the next centerline point.

For the "branch occurrence" type of disconnected vessel reconnection, the following modifications are made to the above reconnection process. First, for the calculation of the D, instead of pointing directly to the endpoint of the candidate branch defined in Eq. \eqref{eq_D}, branch occurrence needs to connect to the corresponding bifurcation of its candidate branch. Since it is difficult to directly obtain the coordinates of the corresponding bifurcation point, we use the minimum distance of all points in the candidate branch as a rough estimate of D, as defined in Eq. \eqref{DPC_new}. Additionally, in the neighborhood set, constraints are applied using historical directions rather than the overall connection vector. Specifically, the offset vectors of neighboring nodes must form angles with the historical forward directions $o_{-1} \text{ and } (o_{-1} + o_{-2})$  that do not exceed  $90^\circ$.
\begin{equation}
    \resizebox{0.52\linewidth}{!}{
    $D({A_k}) =  - \mathop {\min }\limits_{{p_t} \in {\kern 1pt} {\kern 1pt} C{L_j}} {\kern 1pt} {\kern 1pt} {\kern 1pt} {\left\| {{A_k} - {p_t}} \right\|_2}$
   }
\label{DPC_new}
\end{equation}

\subsubsection{Evaluation for the reconnection}
{To avoid incorrect vascular reconnection, it is necessary to evaluate on the reconnection. In the evaluation phase, we input $P$ sequences (generated from the stitched centerline and $CL_{j}$), and compute the sum of mean values of the two $P$ sequences. A predefined threshold (e.g., 1) is then applied to determine the reconnection's correctness. Additionally, we introduce the ADF (Augmented Dickey-Fuller) test to consider the stationarity of $P$ sequence of the stitched centerline and its grayscale value sequence. The p-values from ADF test for both sequences are incorporated as penalty terms and accumulated onto the threshold. To ensure the grayscale continuity of the same blood vessel, the five voxels at the tail of $V_{i}$ and at the head of $CL_{j}$ will also be included in the grayscale value sequence.} 
Reconnections meeting the evaluation criteria are considered correct, allowing the stitched centerlines to proceed to the vessel reconstruction stage. Otherwise, they do not proceed to the next stage, and the "disconnected" vessels are removed from the predicted mask to obtain the refined prediction.

\subsection{Vascular reconstruction stage}
\subsubsection{Implicit neural representation for lumen segmentation}
\begin{figure}[ht]
  \centering
  \includegraphics[width=\linewidth]{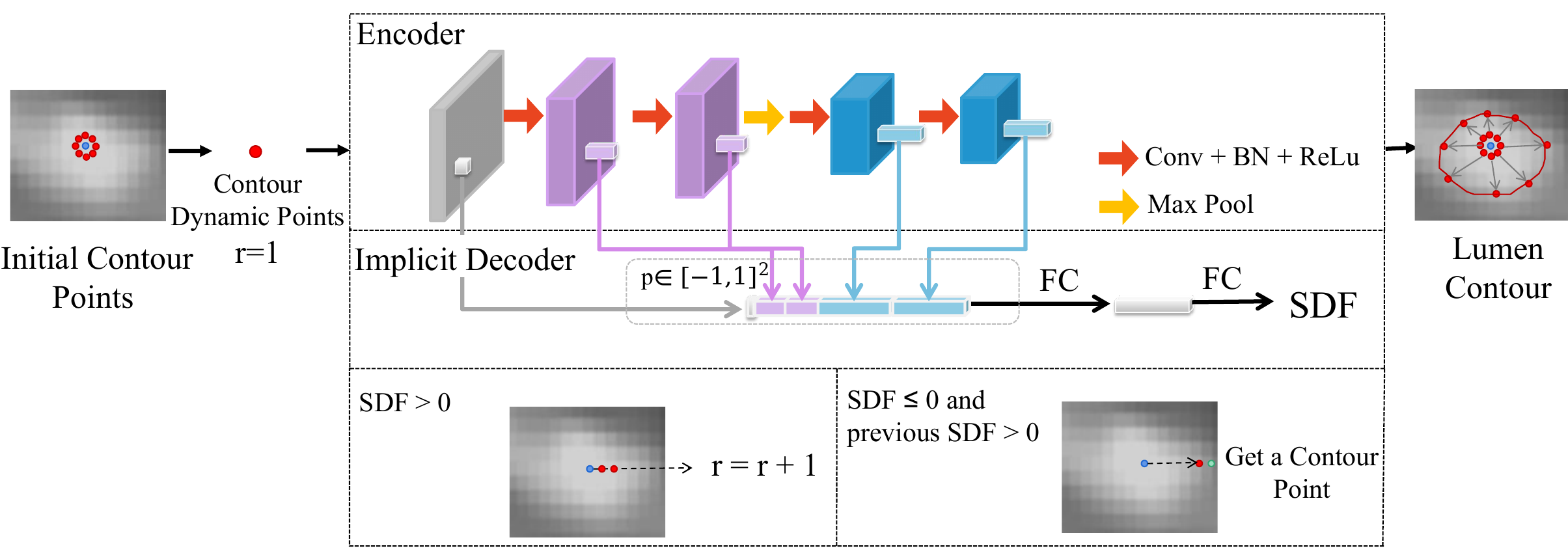}
  \caption{
  Coronary artery luminal contour extraction model based on Implicit neural representation (INR).
  }
  \label{INR1}
  \setlength{\belowcaptionskip}{-1cm}
\end{figure}

Referring to the IOSNet \citep{khan2022implicit} structure used for organs at risk segmentation in head and neck, we devise a luminal contour extraction model based on INR, as illustrated in Fig. \ref{INR1}. This model comprises an encoder and an implicit decoder.

The encoder employs a CNN architecture, taking 2D cross-sections of lumens as input. Following two convolutional layers, a pooling layer, and two additional convolutional layers, the encoder generates a series of feature maps. 
In the implicit decoder, each point $p$ undergoes individual segmentation processing. To achieve this, we extract local and global features associated with point $p$ from the encoder. In the case of 2D sections constructed around centerline points, the coordinates of point $p$ can be normalized to $([-1,1])^2$, allowing for extraction of corresponding coordinates from each layer of the encoder and their connection.
The implicit decoder uses a fully connected neural network to take the extracted features of point 
$p$ as input and output the signed distance value to the luminal boundary, learning a continuous luminal segmentation function.
\begin{equation}
\resizebox{0.45\linewidth}{!}{
$SDF(CS,p) = D(E(CS)_p)$
}
\end{equation}
where $CS$ represents the 2D lumen cross-section, and $E(CS)_p$
represents the features obtained by concatenating the 
$p$-point features extracted from each layer of the encoder. $D(\cdot)$ represents the implicit decoder.

\subsubsection{Training and testing}
To train the network, we construct 2D cross-sections and their segmentation masks along the centerlines of the training set. The signed distance function (SDF) is then computed on these masks. To simplify the calculation, we sample paired samples $(p_{i},sdf_{i})$ from regions near the lumen on the cross-sections. 
Specifically, at each cross-section, we perform multivariate normal distribution sampling with a mean at the center of the cross-section and a variance of 
$\max \left( 10, 5 \times \mathop {\max }\limits_{{p_i} \in {\kern 1pt} {\kern 1pt} CS} (sdf(p_i)) \right)$. 
Points along all lumen boundaries are also sampled for accuracy.

Suppose there are a total of M sampled points across all cross-sections in one iteration of training, and the mean squared error (MSE) is utilized as the supervisory criterion:
\begin{equation}
L = MSE \left( \{ sdf(CS, p_i), sdf_i \}_{i=1}^M \right)
\end{equation} 

Based on the prior that the lumen center is close to the center of the cross-section, we employ a contour point prediction method where contours grow radially outward from the inside. Initially, the 2D cross-section constructed along the central axis is divided into n segments (n > 8) to obtain initial contour points. 
{Each point is then moved outward along its radius and positioned based on the following criteria: 1) \textit{Basic Criterion}: When the point reaches the critical interface where SDF transits from positive to negative, select the position where the absolute SDF value is closer to zero; 2) \textit{Handling Empty Segmentation}: Preventing empty segmentation by maintaining at least the initial positions; 3) \textit{Handling Over-Expansion}: Calculating the weighted mean of SDF values at the center of adjacent cross-sections and itself (denoted as SDF$_{center}$). Then the expansion radius should range from SDF$_{center}$ to $3\times$SDF$_{center}$.}
This process is repeated until all contour points are positioned. Finally, these points are sequentially fitted into a closed curve to obtain the lumen contour.
\subsubsection{Implicit surface reconstruction}

\begin{figure}[ht]
  \centering
  \includegraphics[width=\linewidth]{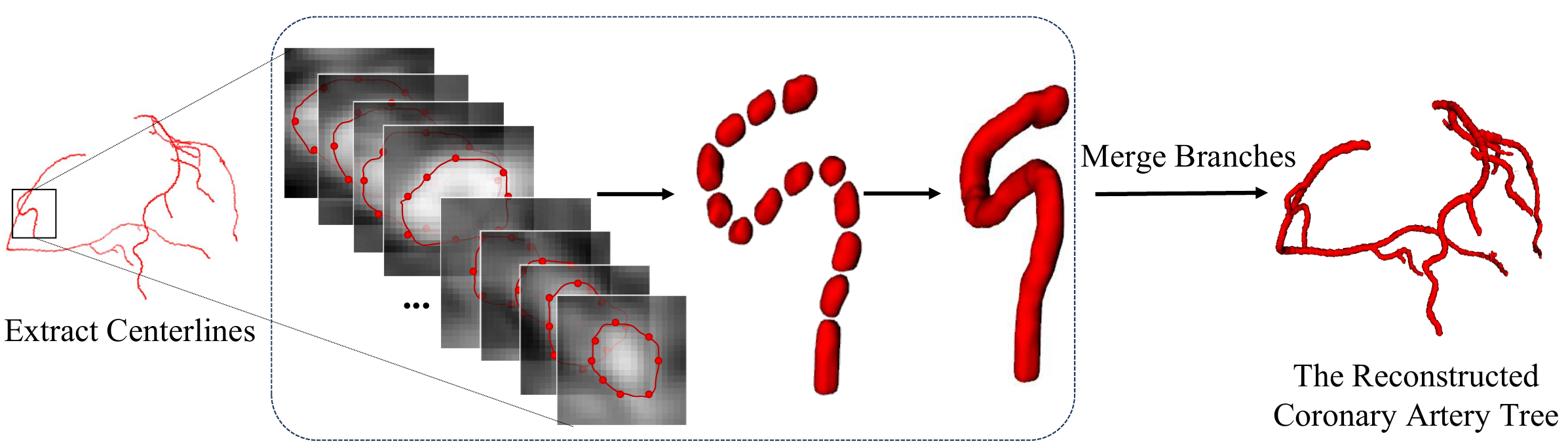}
  \caption{
  The process of coronary artery reconstruction based on implicit extrusion surfaces (IES).
  }
  \label{rec}
  \setlength{\belowcaptionskip}{-1cm}
\end{figure}
The vascular tree reconstruction is illustrated in Fig. \ref{rec}. Initially, a representation of the cross-sectional contours will be employed using the Partial Shape-Preserving Spline method \citep{li2011partial}. This method constructs smooth and continuous curves from discrete boundary points of the cross-sectional contours to achieve precise representation of irregular cross-sectional profiles. Subsequently, an implicit tubular model representation based on implicit extrusion surfaces (IES) \citep{HONG2020} will be utilized to blend the individual sectional models, thereby constructing an implicit tubular model with varying cross-sectional profiles. In the implicit tubular model representation based on IES, it is necessary to first construct two perpendicular implicit surfaces 
$F_{1}(x,y,z)=0$ and $F_{2}(x,y,z)=0$ based on the centerline. Their intersection forms an implicitly fitted curve, representing the stretching path. Subsequently, along this intersection, the 2D implicit curves representing the cross-sections can be extruded into 3D implicit surfaces, namely implicit extrusion surfaces. 
For different cross-sections $C_{i}(x,y)$, a weighted summation using 1D PSPS \citep{li2011partial} functions $B_{i}(t)$ along the given centerline is required to construct a tubular structure model with varying cross-sectional profiles. The overall implicit function expression of the model is as follows:
\begin{equation}
\resizebox{0.88\linewidth}{!}{
$ F(x,y,z) = \sum_{i=1}^{L} C_i(F_{1}(x,y,z),F_{2}(x,y,z)) \cdot B_i(t) = 0 $
}
\end{equation} 

\section{Experiments and analysis}
\subsection{Datasets}
We validated our framework on two datasets, one public (ASOCA) and one in-house (PDSCA). \textbf{ASOCA}:  This public dataset from the MICCAI 2020 Automated Segmentation of Coronary Arteries challenge includes 40 training cases (20 normal, 20 diseased) with voxel labels manually annotated by three experts. Additionally, there are 20 CTA test images. Both images and annotations have anisotropic resolution, with Z-axis resolution of 0.625mm and in-plane resolution of 0.3-0.4mm. \textbf{PDSCA}: The PDSCA dataset is a private dataset collected from Peking Union Medical College Hospital, and the study has been approved by its ethics review committee. The PDSCA includes 100 labeled CTAs with a resolution of 0.3×0.3×0.5$mm^3$ and is annotated by three radiologists.
Both ASOCA and PDSCA are 512×512×$S_z$, with $S_z$ between 155 and 353. For ASOCA and PDSCA, the same  five-fold cross-validation method as {\citep{10265156}} is used for evaluation.
\subsection{Experimental setup}
In vascular segmentation stage, the network parameters were set consistent with nnUNetTrainerV2, and the model was trained and tested on an NVIDIA GeForce GTX 1080Ti using Python 3.7.11 and PyTorch 1.7.1. In vascular reconnection stage, 
{ CascadeForestClassifier \citep{zhou2019forest} implemented in
Python deepforest library was trained as the centerline classifier. During the training of CFC, samples were taken in the centerline region, the non-centerline region in the lumen, and also the area outside the lumen with a distance of no more than 7 from the tube wall. All centerline points were selected, with the number of selected centerline points as N. Additionally, 2N points were sampled from each of the other two regions. Overall, the ratio of positive to negative samples was controlled at about 1:4. }


\subsection{Evaluation metrics}
For segmentation masks, the Dice coefficient and 95 percent Hausdorff distance (HD95) are used for evaluation. For the centerline in the reconnection stage, the ability to track all blood vessels is measured using Overlap (OV). Besides, new metrics are proposed for the evaluation of stitched and broken centerlines -- Reconnection Accuracy, Reconnection Sensitivity, and Reconnection Specificity, denoted by RecAcc, RecSen, and RecSpe, defined as: 


\begin{eqnarray}
   & RecAcc = \frac{(TP_{b} + TP_{s}) + TN_{b}}{(TP_{b} + TP_{s}) + TN_{b} + (FP_{b} + FP_{s}) + FN_{b}}  \\
   & RecSen = \frac{TP_{b} + TP_{s}}{(TP_{b} + TP_{s}) + FN_{b}} \label{3.2-1}  \\
   & RecSpe = \frac{TN_{b}}{TN_{b} + (FP_{b} + FP_{s})} \label{3.2-2} 
\end{eqnarray}
where subscript $_b$ and subscript $_s$ represent broken centerlines and stitched centerlines respectively. The labels of TP, TN, FP and FN are assigned by comparing the two centerlines with the ground truth. RecAcc evaluation includes the overall accuracy of connection and removal in cases, and RecSen reflects the true "connection". And RecSpe reflects the true "removal".
\subsection{Comparison with state-of-the-art methods}
We conducted quantitative and qualitative comparison of our proposed CorSegRec with several deep learning-based segmentation methods on the ASOCA and PDSCA datasets using five-fold cross-validation. 
{To comprehensively evaluate the performance of our proposed method, the baseline methods include the benchmark segmentation network, i.e., ResUNet \citep{kerfoot2019left}, as well as seminal techniques specialized for coronary artery segmentation, i.e., MPSPNet \citep{zhu2021coronary}, 3D-FFR-Unet \citep{song2022automatic}, and ADE-HTL Net \citep{10265156}. In addition, we also compared with some advanced topology-based deep learning methods for tubular structure extraction, including TopNet \citep{keshwani2020topnet}, CS$^{2}$-Net \citep{mou2021cs2}, and clDice \citep{shit2021cldice}.}

Table \ref{ASOCA} presents the results on the ASOCA dataset. Compared to the classical segmentation network ResUNet \citep{kerfoot2019left}, CorSegRec achieves improvement of 6.50\% in Dice and 6.44mm in HD95, respectively. Among other methods for tubular structure extraction, ADE-HTL Net \citep{10265156} performs the best. It consists of an anatomy-dependent encoding module and a multi-task ResUNet for learning coronary artery hierarchical topology, providing a specialized solution for anatomy and topology preservation. In contrast, CorSegRec does not require obtaining a region of interest beforehand. Instead, it directly utilizes the spatial distribution of "disconnected" vessels and local vessel information from the segmentation results of nnU-Net as priors, resulting in further improvement of 2.45\% in Dice and 4.18mm in HD95.
Table \ref{PDSCA} presents quantitative comparison with other methods on the PDSCA dataset. From the results, CorSegRec outperforms ResUNet \citep{kerfoot2019left} by 14.82\% in Dice and 7.70mm in HD95, showing even more significant improvement compared to those on the ASOCA validation set. Compared to other methods, CorSegRec still demonstrates more satisfactory segmentation performance.
\begin{table}[ht!]
  \small
  \centering
  \caption{
  {{Performance comparison of Dice and HD95 between our method (CorSegRec) and other state-of-the-art methods on ASOCA. (mean $\pm$ std, with the best results shown in {\bfseries bold})}}}
  \resizebox{0.84\columnwidth}{!}{%
    \begin{tabular}{l|cc}
    \toprule[1pt]
    Method & Dice(\%) $\uparrow$ & HD95(mm) $\downarrow$  \\
    \midrule
    ResUNet\citep{kerfoot2019left} & 82.03 $\pm$ 1.39 & 7.51 $\pm$ 0.85   \\
    MPSPNet\citep{zhu2021coronary} & 83.08 $\pm$ 1.46 & 6.91 $\pm$ 0.62 \\
    CS$^{2}$-Net\citep{mou2021cs2} & 82.80 $\pm$ 1.51 & 7.11 $\pm$ 0.68 \\
    TopNet\citep{keshwani2020topnet} & 82.54 $\pm$ 1.03 & 6.03 $\pm$ 0.44 \\
    clDice\citep{shit2021cldice} & 83.57 $\pm$ 1.25 & 6.64 $\pm$ 0.41 \\
    3D-FFR-Unet\citep{song2022automatic} & 83.85 $\pm$ 1.30 & 6.77 $\pm$ 0.52 \\
    ADE-HTL Net {\citep{10265156}} & 86.08 $\pm$ 1.21 & 5.25 $\pm$ 0.43 \\
    \textbf{Ours (CorSegRec)} & \textbf{88.53} $\pm$ \textbf{1.81} & \textbf{1.07} $\pm$ \textbf{0.60} \\
    \bottomrule[1pt]
    \end{tabular}%
    }
  \label{ASOCA}%
  \vspace{-2mm}
\end{table}

\begin{table}[ht!]
  \small
  \centering
  \caption{
  {{Performance comparison of Dice and HD95 between our method (CorSegRec) and other state-of-the-art methods on PDSCA. (mean $\pm$ std, with the best results shown in {\bfseries bold})}}}
  \resizebox{0.84\columnwidth}{!}{%
    \begin{tabular}{l|cc}
    \toprule[1pt]
    Method & Dice(\%) $\uparrow$ & HD95(mm) $\downarrow$  \\
    \midrule
    ResUNet\citep{kerfoot2019left} & 70.25 $\pm$ 1.46 & 9.33 $\pm$ 0.22   \\
    MPSPNet\citep{zhu2021coronary} & 73.60 $\pm$ 1.17 & 7.08 $\pm$ 0.24 \\
    CS$^{2}$-Net\citep{mou2021cs2} & 76.53 $\pm$ 1.29 & 7.33 $\pm$ 0.31 \\
    TopNet\citep{keshwani2020topnet} & 76.65 $\pm$ 1.02 & 6.35 $\pm$ 0.34 \\
    clDice\citep{shit2021cldice} & 76.72 $\pm$ 0.86 & 6.29 $\pm$ 0.18 \\
    3D-FFR-Unet\citep{song2022automatic} & 79.35 $\pm$ 1.04 & 6.19 $\pm$ 0.50 \\
    ADE-HTL Net {\citep{10265156}} & 82.14 $\pm$ 1.15 & 5.68 $\pm$ 0.23 \\
    \textbf{Ours (CorSegRec)} & \textbf{85.07} $\pm$ \textbf{0.79} & \textbf{1.63} $\pm$ \textbf{0.55} \\
    \bottomrule[1pt]
    \end{tabular}%
    }
  \label{PDSCA}%
\end{table}

\begin{figure*}[htbp]
\setlength{\abovecaptionskip}{-1mm}
\centering 
\includegraphics[width=\textwidth]{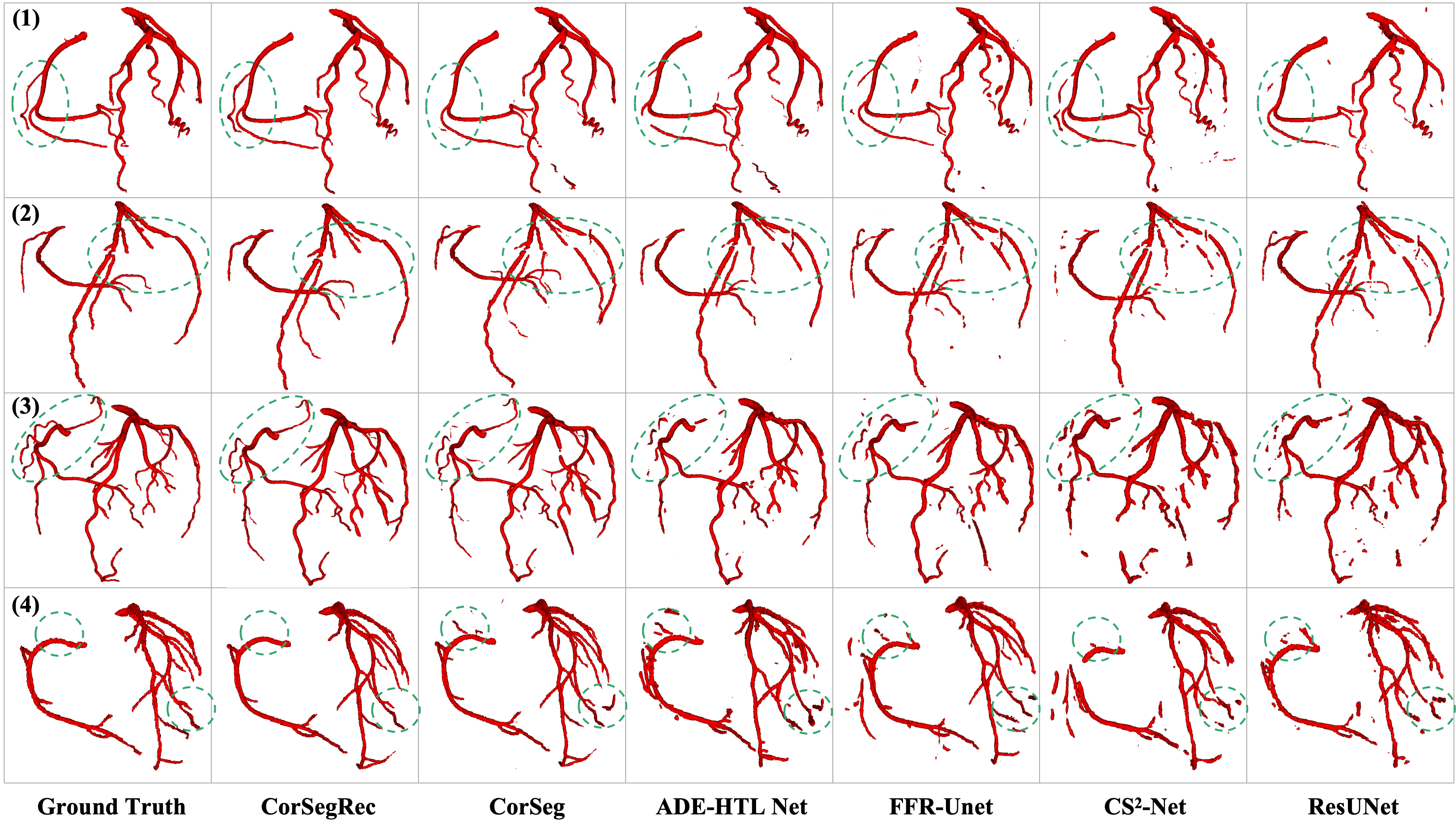}
\caption{The visual results of CorSegRec compared to other methods on the ASOCA and PDSCA datasets. Cases (1) and (2) are from ASOCA, while Cases (3) and (4) are from PDSCA. The third column displays the results of the CorSeg method at the segmentation stage of our approach, and the second column shows the final results of CorSegRec obtained after the reconstruction stage of our approach. 
As indicated by the green dashed circles, our method can achieve topologically and visually
similar structures to the ground truth, whereas other methods generally result in disconnected and incomplete vessel segments.}
\label{copare}
\vspace{-3mm}
\end{figure*}

The visualization results on the ASOCA and PDSCA datasets are presented in Fig. \ref{copare}. Cases (1) and (2) are from ASOCA, while Cases (3) and (4) are from PDSCA. In Case (1), short disconnected small vessels, indicated by the green dashed circles, are distant from the coronary tree. After reconnection and reconstruction, topologically and visually similar structures to the ground truth are restored, whereas in other methods, this segment of vessels remains incomplete. Case (2) demonstrates the ability of CorSegRec to remove neighboring non-coronary structures, which bears significant resemblance to the coronary tree in direction and position. In the PDSCA annotations, the vessels are more complete, and due to smaller spacing, they appear thicker visually compared to ASOCA. In Case (3), CorSegRec successfully reconnects one disconnected vessel to the coronary tree, but another vessel with fewer predicted voxels is incorrectly removed. In Case (4), CorSegRec accurately removes non-coronary tubular structures, while other methods still exhibit numerous false positive vessels, confirming the necessity of removing nearby unrelated structures. In CorSegRec, these false positive vessels are successfully eliminated through multiple rounds of filtering and reconnection. 
\subsection{Ablation study}
To demonstrate the effectiveness of the components proposed in CorSegRec, we conduct ablation experiments on key structures within CorSegRec, including the loss function in the segmentation stage, the patch size of CFC, the probability term weight in the DPC random walk algorithm, the three components D, P, and C in DPC, the two-level neighbor sets, the normalized P, the branch-occurrence reconnection, the vascular reconnection stage, per cross-section sampling points and per cross-section contour moving points.

\subsubsection{Effectiveness of NSDT soft-clDice loss}
\begin{table}[ht!]
\small
  \centering
\caption{Ablation experiment on the loss functions in the vascular segmentation stage.}
  \resizebox{0.9\columnwidth}{!}{%
\begin{tabular}{c|cc|cc}
\toprule[1pt]   
\multirow{2}{*}{Dataset} & \multicolumn{2}{c|}{Loss Function} & \multirow{2}{*}{Dice (\%) $\uparrow$} & \multirow{2}{*}{HD95 (mm) $\downarrow$} \\
\cline{2-3}
 & soft-clDice & NSDT soft-clDice \\
\hline
\multirow{2}{*}{ASOCA}  & \checkmark & & 85.66 & 5.22 \\
  & & \checkmark & \textbf{87.13} & \textbf{5.06} \\
\multirow{2}{*}{PDSCA}  & \checkmark & & 82.22 & 12.78 \\
  & & \checkmark & \textbf{83.37} & \textbf{9.91} \\
\bottomrule[1pt]
\end{tabular}
}
\label{loss}
\end{table}

Table \ref{loss} displays the results of ablation experiments on the loss functions. On the ASOCA test set, our NSDT soft-clDice outperformed soft-clDice in terms of Dice coefficient and HD95. Similarly, on the PDSCA dataset, 
NSDT soft-clDice 
also demonstrated improvement in both metrics when compared with soft-clDice. It can be observed in experiments that if coronary artery CTA is resampled to the same spacing, segmentation performance deteriorates due to reduced resolution. Therefore, original-spacing CTA datasets were used in our experiments. For coronary datasets with two different spacings, the combination of NSDT soft-clDice and Dice exhibited relatively stable segmentation performance.
\begin{table}[htbp]
  \centering
\caption{Ablation experiment on cascade forest classifier patch size.}
\resizebox{0.85\columnwidth}{!}{
\begin{tabular}{c|c|ccc}
\toprule[1pt]
Dataset & Size & Acc (\%) & Sen (\%) & Spe (\%) \\
\hline   
 & 15$\times$15$\times$15 & 94.02 & 87.20 & 95.73 \\
ASOCA  & 11$\times$11$\times$11 & 94.41 & 88.85 & \textbf{95.80} \\
 & 7$\times$7$\times$7 & 94.37 & 88.75 & 95.78 \\
 & Multiscale Patch (15,7) & \textbf{94.48} & \textbf{89.53} & 95.72 \\

 & 15$\times$15$\times$15 & 91.44 & 77.10 & 95.01 \\
PDSCA & 11$\times$11$\times$11 & \textbf{92.17 }& 78.55 & 95.54 \\
 & 7$\times$7$\times$7 & 92.07 & 77.87 & \textbf{95.59} \\
 & Multiscale Patch (15,7) & 91.91 & \textbf{79.58} & 94.97 \\
\bottomrule[1pt]
\end{tabular}
}
\label{pz}
\end{table}
\vspace{-1mm}
\subsubsection{Impact of patch size for cascade forest classifier}
We tested the impact of using a single size patch and a fusion of two patch sizes on the cascade forest classifier, with results shown in Table \ref{pz}. On both datasets, using a single size of $11\times11\times11$ outperforms the larger size of $15\times15\times15$ and the smaller size of $7\times7\times7$ in all three metrics. This size could encompass the largest vascular lumen area while minimizing inclusion of surrounding vascular structures. Considering the wide range of coronary artery diameters, we concatenated inputs to the classifier from multiple sizes of patches, with the larger patches downsampled to the same size as the smaller ones. Multiscale patches constructed using side lengths of 15 and 7 achieve higher accuracy and sensitivity than single-size patches on ASOCA and higher sensitivity on PDSCA, indicating that multi-scale patches are more effective in identifying true centerline points.
\subsubsection{Impact of parameter $\omega$}
For the D, P, and C terms in the DPC walk algorithm, the variation magnitude of D and C is similar among the same-level neighboring points. To simplify the parameter tuning process, the weights of D and C are both set to 1, while P is weighted using $\omega$. 
Ablation experiments on $\omega$ were conducted as shown in Table \ref{omega}, with $\omega$ ranging from 0 to 7 in step of 1, evaluated using RecAcc. On both the ASOCA and PDSCA datasets, when $\omega$ is set to 0, indicating that the random walk algorithm does not use the probability term, the reconnection accuracy is the lowest. As $\omega$ increases to 5, RecAcc gradually rises to its maximum value before slowly decreasing. This is because when the probability term becomes too dominant, the influence of distance (D) and cosine similarity (C) diminishes, making the algorithm trapped in local high-probability areas and deviate from the correct centerline.
\begin{table}[htbp]
\caption{Ablation experiment on parameter $\omega$ of the DPC walk algorithm (evaluated using RecAcc).}
\resizebox{0.94\columnwidth}{!}{
\begin{tabular}{c|cccccccc}
\toprule[1pt]     
Dataset  &  0  & 1  & 2  & 3  & 4  & 5 & 6  & 7\\

\hline
ASOCA & 52.12 & 73.21 & 83.78 & 87.08 & 89.72 & \textbf{92.22} & 92.22 & 90.14 \\
PDSCA & 65.92 & 79.49 & 80.20 & 79.05 & 81.92 & \textbf{82.25} & 81.98 & 81.44 \\
\bottomrule[1pt]
\end{tabular}
}
\label{omega}
\end{table}

\begin{table}[htbp]
\caption{Ablation experiment on components of the DPC walk algorithm.}
\resizebox{\columnwidth}{!}{
\begin{tabular}{c|c|ccc|c}
\toprule[1pt]
Dataset & Method & RecAcc(\%) & RecSen(\%) & RecSpe(\%) & OV(\%) \\
\hline
 \multirow{5}{*} {ASOCA}& CorSegRec w/o DPC & - & - & - & 86.27 \\
 & CorSegRec w/ DPC & \textbf{92.22} & \textbf{98.20} & 82.79 & \textbf{90.14} \\
 & CorSegRec w/ DP & 82.02 & 84.60 & 78.28 & 88.84 \\
 & CorSegRec w/ PC & 83.49 & 84.40 & 82.18 & 89.31 \\
 & CorSegRec w/ DC & 50.12 & 14.49 & \textbf{90.00} & 85.12 \\
 
\multirow{5}{*} {PDSCA}& CorSegRec w/o DPC & - & - & - & 85.26 \\
 & CorSegRec w/ DPC & \textbf{82.25} & \textbf{89.28} & 76.81 & \textbf{89.18} \\
  & CorSegRec w/ DP & 74.21 & 80.81 & 68.89 & 88.84 \\
 & CorSegRec w/ PC & 76.23 & 63.93 & 86.27 & 88.58 \\
 & CorSegRec w/ DC & 63.33 & 18.57 & \textbf{93.59} & 86.63 \\
\bottomrule[1pt]
\end{tabular}
\label{DPC}
}
\end{table}


\subsubsection{Effectiveness of DPC components}
We conducted ablation experiments on the components D, P, and C of the DPC walk algorithm, evaluating the disconnection and reconnection of the centerline during the vascular reconnection process using custom metrics RecAcc, RecSen, and RecSpe, and overall centerline coverage using OV. The results are shown in Table \ref{DPC}. By combining D, P, and C, subsets of DPC are obtained: $ \emptyset$, DPC, DP, PC, and DC. Among these subsets, the DPC walk performs better than others in terms of RecAcc, RecSen, and OV metrics. Since the focus of the algorithm is on connection rather than removal, the specificity (RecSpe) is relatively lower for DPC compared to other subsets. The subset without P—DC walk—performed the worst, indicating that P is a key component of the DPC random walk algorithm. PC and DP rank similarly within the subsets because they both include the probability term. However, PC walk lacks guidance to the destination, which poses difficulty in long-distance reconnections, while the DP algorithm experiences local fluctuations and deviates from the true centerline when the probability term is inaccurate.

\subsubsection{Effectiveness of second-level neighbor sets}

To explore the impact of different levels of neighbor sets on model performance, we compared the second-level neighbor set with the first-level neighbor set. As shown in Table \ref{Neighbor}, the RecAcc has been improved by 5.94\% on the ASOCA dataset and by 2.79\% on the PDSCA dataset, respectively. This improvement is attributed to the ability of the second-level neighbor set to establish reasonable connections during the reconnection process, especially between points that are distant but potentially connected. This capability is crucial for maintaining connectivity paths from the head $CL_{j}$ to the tail $V_{i}$, which may span long distances or complex structures.
\begin{table}[htbp]
\centering
\caption{Ablation experiment on different levels' neighbor sets.}
\resizebox{0.95\columnwidth}{!}{
\begin{tabular}{c|c|ccc|c}
\toprule[1pt]
Dataset & Neighbor Set & RecAcc(\%) & RecSen(\%) & RecSpe(\%) & OV(\%) \\
\hline
\multirow{2}{*} {ASOCA} & first level & 86.28 & 91.05 & 78.76 & 89.64 \\
 & second level & \textbf{92.22} & \textbf{98.20} & \textbf{82.79} & \textbf{90.14} \\
\multirow{2}{*} {PDSCA}  & first level & 79.46 & 85.11 & 74.03 & 88.83 \\
 & second level & \textbf{82.25} & \textbf{89.28} & \textbf{76.81} & \textbf{89.18} \\
\bottomrule[1pt]
\end{tabular}

}
\label{Neighbor}
\end{table}

\begin{table}[ht!]
\small
  \centering
\caption{Ablation experiment on branch-occurrence reconnection.}
  \resizebox{\columnwidth}{!}{%
\begin{tabular}{c|ccc|ccc|c}
\toprule[1pt]   
\multirow{2}{*}{Dataset} & \multicolumn{3}{c|}{Reconnection Type} & \multirow{2}{*}{RecAcc (\%) } & \multirow{2}{*}{RecSen (\%) } & \multirow{2}{*}{RecSpe (\%) } & \multirow{2}{*}{ OV(\%)} \\
\cline{2-4}
& 1 & 2 & 3 (B-Occ) \\
\hline
 \multirow{2}{*} {ASOCA} & \checkmark & \checkmark & & 81.59 & 80.75 & 82.79 & 89.07 \\
& \checkmark & \checkmark & \checkmark  & \textbf{92.22} & \textbf{98.20} & \textbf{82.79} & \textbf{90.14} \\
\multirow{2}{*} {PDSCA} & \checkmark & \checkmark  &  & 79.98  & 82.16 & \textbf{80.27} & 88.89\\
 & \checkmark & \checkmark & \checkmark & \textbf{82.25} & \textbf{89.28} & 76.81 & \textbf{89.18}\\
    \bottomrule[1pt]
\end{tabular}
}
\label{branch_occurrence}
\end{table}
\subsubsection{Effectiveness of branch-occurrence reconnection}

We conducted an ablation experiment to explore the impact of branch occurrence reconnection on model performance. As shown in Table \ref{branch_occurrence}, reconnecting at branch occurrences has improved the RecAcc by 10.63\% on the ASOCA dataset and by 2.27\% on the PDSCA dataset, respectively. This improvement stems from the ability of branch occurrence reconnection to maintain vascular network connectivity in complex coronary artery topologies. This means that even in cases where vascular branches are distant but potentially connected, the reconnection algorithm can effectively establish reasonable connections, thus preventing vascular discontinuity or overlapping path issues.

\subsubsection{Effectiveness of vascular reconnection \& reconstruction stages}
To explore the impact of the vascular reconnection \& reconstruction stages on model performance, we conducted the ablation study as shown in Table \ref{reconnection_stage}. Introducing the vascular reconnection \& reconstruction stages has improved the Dice by 1.4\% on the ASOCA dataset and by 1.7\% on the PDSCA dataset, indicating that reconnecting disconnected vascular segments enhances the model's performance. This improvement is primarily attributed to vascular reconnecting optimizing the topology of the vascular network, thereby enhancing the model's ability to accurately model complex vascular pathways.

\begin{table}[ht!]
\small
  \centering
\caption{Ablation experiment on vascular reconnection \& reconstruction stages.}
  \resizebox{0.7\columnwidth}{!}{%
\begin{tabular}{c|cc|cc}
\toprule[1pt]   
\multirow{2}{*}{Dataset} & \multicolumn{2}{c|}{Stage} & \multirow{2}{*}{Dice (\%) $\uparrow$} & \multirow{2}{*}{HD95 (mm) $\downarrow$} \\
\cline{2-3}
& Stage 1 & Stage 2\&3 \\
\hline
\multirow{2}{*}{ASOCA} & \checkmark  & & 87.13 & 5.06 \\
& \checkmark & \checkmark  & \textbf{88.53} & \textbf{1.07} \\
\multirow{2}{*}{PDSCA} & \checkmark &  & 83.37 & 9.91 \\
 & \checkmark & \checkmark  & \textbf{85.07} & \textbf{1.63}  \\
\bottomrule[1pt]
\end{tabular}
}
\label{reconnection_stage}
\end{table}

\subsubsection{Impact of per cross-section sampling points} Table \ref{traninr} presents the results of ablation experiments conducted during INR training, focusing on the number of sampled points per cross-section. From the results, it is evident that as the number of sampled points per cross-section increases, performance gradually improves. However, the rate of improvement slows down until it eventually converges. Notably, optimal performance is achieved for both datasets when the number of sampled points is set to $15 \times \text{max}(sdf(CS))$. Consequently, these two parameter settings were selected for further testing.

\begin{table}[htbp]
\centering
\caption{Ablation experiment on the number of sampled points per cross-section during INR training. \textit{Num} represents the sampling number per cross-section. \textit{Time} represents the average iteration time.}
\resizebox{0.75\columnwidth}{!}{
\begin{tabular}{c|cc|c}
\toprule[1pt]
\multirow{2}{*}{\textit{Num}} & \multicolumn{2}{c|}{Dice (\%) $\uparrow$} & \multirow{2}{*}{Time (s) $\downarrow$} \\
    \cline{2-3}
          & ASOCA & PDSCA  \\
\hline
20 & 83.48 & 79.46 & 9.64 \\
30 & 84.33 & 81.47 & 10.14  \\
40 & 84.72 & 81.81 & 13.49 \\
50 & 85.13 & 82.37 & 13.66  \\
60 & 86.19 & 82.68 & 14.29 \\
$5 \times \max(\text{sdf}({CS}))$ & 85.87 & 82.04 & 8.20  \\
$10 \times \max(\text{sdf}({CS}))$ & 86.04 & 83.08 & 11.21 \\
$15 \times \max(\text{sdf}({CS}))$ & \textbf{86.63} & \textbf{83.20} & 15.84  \\
\bottomrule[1pt]
\end{tabular}
}
\label{traninr}
\end{table}


\subsubsection{Impact of per cross-section contour moving points}

We conducted ablation experiments on the number of contour points during testing, where we set seven different values, as shown in Table \ref{testinr}. On the ASOCA dataset, we observed optimal performance when the number of contour points was set to 20. This suggests that having a higher number of contour points contributes to better performance on the ASOCA dataset. Conversely, on the PDSCA dataset, we found that performance was best when the number of contour points was set to 12.

\begin{table}[htbp]
\centering
\caption{Ablation experiment on the number of dynamic contour points during INR testing. \textit{Num} represents the number of contour dynamic points. \textit{Time} represents the average inference time.}
\resizebox{0.95\columnwidth}{!}{
\begin{tabular}{c|cc|c}
\toprule[1pt]
\multirow{2}{*}{\textit{Num}}  & \multicolumn{2}{c|}{Dice (\%) $\uparrow$} & \multirow{2}{*}{Time (s) $\downarrow$} \\
    \cline{2-3}
          & ASOCA & PDSCA  \\
\hline
8 & 85.21 & 83.12 & 0.15 \\ 
12 & 86.28 & \textbf{83.20} & 0.23  \\
16 & 86.44 & 82.95 & 0.31 \\
20 & \textbf{86.63} & 83.00 & 0.38\\
24 & 86.45 & 83.00 & 0.47\\
$ \min(\max(4\times \text{sdf}({CS}_{\text{center}}), 8), 16)$ & 86.33 & 83.04 & 0.18  \\
$ \min(\max(4\times \text{sdf}({CS}_{\text{center}}), 8), 24)$ & 86.42 & 83.03 & 0.20 \\
\bottomrule[1pt]
\end{tabular}
}
\label{testinr}
\end{table}


{Finally, we summarize the key hyperparameter settings involved in the ablation study and their optimized values in Table \ref{parameters}.}
\begin{table}[htbp]
  \centering
\caption{Summary of key hyperparameter settings.}
  \resizebox{1.0\columnwidth}{!}{%
\begin{tabular}{c|c|c}
\toprule[1pt]   
Stage & rameter & Final set value \\
\cline{2-3}
\hline
\multirow{2}{*}{Reconnection} & Patch size  & Multiscale Patch (15,7) \\
& $\omega$ (weight of $P$) & 5 \\
\multirow{2}{*} {Reconstruction} & Sampling points per cross-section & $15 \times max(sdf(CS))$ \\
 & Contour points per cross-section & ASOCA: 20, PDSCA: 12  \\
    \bottomrule[1pt]
\end{tabular}
}
\label{parameters}
\end{table}

\section{Conclusion}
\label{sec:guidelines}

In this paper, we proposed CorSegRec, a three-stage coronary artery extraction protocol consisting of vascular segmentation, vascular reconnection, and vascular reconstruction. To address the issue of inaccurate topology in neural network segmentation results, we designed a centerline Dice loss function combined with centerline distance transformation during the vascular segmentation stage. This approach enhances the connectivity of segmentation results and detects more coronary vessels. For discontinuous vascular segments resulting from automatic coronary segmentation, the DPC Walk algorithm was proposed to reconnect disconnected blood vessels into a complete coronary vascular tree. The algorithm searches for the next centerline point in the neighborhood set, using spatial distribution information (D, C) and local vessel information (P) to connect disconnected blood vessels to the ends of candidate vessels or to form new branches, thus achieving a fully-connected coronary centerline. Additionally, we incorporated INR and implicit modeling to complete the missing vascular structure along the centerline. Compared to end-to-end neural network models, CorSegRec took into account coronary artery anatomy and location characteristics as well as integrating both global and local information for coronary artery segmentation. Extensive experiments on two CTA datasets showed that our CorSegRec outperformed other tubular structure extraction methods, achieving superior volumetric scores and higher vascular connectivity. In addition, adequate ablation studies have been conducted to demonstrate the effectiveness of the proposed method. {It should be noted that the time complexity of CorSegRec is higher than that of other state-of-art methods, due to use of two stages of reconnection and reconstruction. In the future, we will further improve reconnection efficiency, and will also apply CorSegRec to other vascular structures.}

\vspace{-3mm}
\section{Declaration of competing interest}
The authors declare that they have no known competing financial interests or personal relationships that could have appeared to
influence the work reported in this paper.
\vspace{-3mm}
\section{Data availability}
Data will be made available on request.
\section{Acknowledgments}
This work was supported in part by the National Natural Science Foundation of China (grant numbers 62471418, 82441023, U23A20295, 62131015), the Beijing Natural Science Foundation (grant number Z210013), the Key R\&D Program of Guangdong Province, China (grant numbers 2023B0303040001, 2021B0101420006), and the Fujian Provincial Natural Science Foundation of China (grant number 2024J01058).


\printcredits

\bibliographystyle{cas-model2-names}

\bibliography{main}





\end{document}